# Millikelvin temperature cryo-CMOS multiplexer for scalable quantum device characterisation

Anton Potočnik,[1,*] Steven Brebels,[1] Jeroen Verjauw,[1,2] Rohith Acharya,[1,3] Alexander Grill,[1,3] Danny Wan,[1] Massimo Mongillo,[1] Ruoyu Li,[1] Tsvetan Ivanov,[1] Steven Van Winckel,[1] Fahd A. Mohiyaddin,[1] Bogdan Govoreanu,[1] Jan Craninckx,[1] and I. P. Radu[1]

[1] Imec, Kapeldreef 75, 3001 Leuven, Belgium

[2] KU Leuven, Dept. of Materials Engineering (MTM), Kasteelpark Arenberg 44, 3001 Leuven, Belgium

[3] KU Leuven, Dept. of Electrical Engineering (ESAT), Kasteelpark Arenberg 10, 3001 Leuven, Belgium

* Corresponding author, anton.potocnik@imec.be

## Abstract

Quantum computers based on solid state qubits have been a subject of rapid development in recent years. In current Noisy Intermediate-Scale Quantum (NISQ) technology, each quantum device is controlled and characterised through a dedicated signal line between room temperature and base temperature of a dilution refrigerator. This approach is not scalable and is currently limiting the development of large-scale quantum system integration and quantum device characterisation. Here we demonstrate a custom designed cryo-CMOS multiplexer operating at 32 mK. The multiplexer exhibits excellent microwave properties up to 10 GHz at room and millikelvin temperatures. We have increased the characterisation throughput with the multiplexer by measuring four high-quality factor superconducting resonators using a single input and output line in a dilution refrigerator. Our work lays the foundation for large-scale microwave quantum device characterisation and has the perspective to address the wiring problem of future large-scale quantum computers.

# 1. Introduction

Quantum computing technology is a subject of intense research and development. Numerous groups have already demonstrated quantum processors containing several tens of quantum bits (qubits) with unprecedented gate fidelities [1–4]. A path forward is being explored both in the near-term noisy intermediate-scale quantum computer (NISQ) direction with up to a few hundreds of qubits on a chip [5,6], as well as in the direction of a fully fault-tolerant quantum computer with projected millions of qubits [7,8].

To scale up quantum processors, several technological challenges still need to be solved in terms of device fabrication [9–11], high-performance control electronics [12,13] and system integration [14–16]. For example, current implementations require each qubit to have at least one control line between room temperature electronics and the base plate in a dilution refrigerator [14,16]. This is not a scalable strategy for devices containing thousands of qubits, due to spatial limitations and the limited cooling power of dilution refrigerators [17].

Furthermore, state-of-the-art quantum devices exhibit large parameter variations [10,11]. In frequency crowded multiqubit systems, this can lead to an accidental spectral overlap and consequently reduce quantum gate and process fidelities [18]. To improve device reproducibility, the fabrication process is migrating from laboratory environments to industrial-level 200 mm/300 mm fabrication facilities [10]. Equipment developed for mature semiconductor technologies in these facilities offer low variability, near-atomic scale precision and production of a large number of devices, enabling statistical-level electrical characterisation [19]. The latter provides a valuable insight into the material properties at room and cryogenic temperatures, as well as enables further development of fabrication processes [19,20]. However, massive characterisation of quantum devices at cryogenic temperatures is limited by a non-scalable number of required input and output lines in a dilution refrigerator [17].

To alleviate the wiring problem and increase characterisation throughput, a multiplexing scheme is needed at the base temperature of the dilution refrigerator (figure 1(a)) [17]. In such a scheme, a user interface and classical computing/control platform communicates with the control and readout electronics using minimal number of lines. Control electronics based on cryo-CMOS technology can be placed at the 1-4 K stage in the refrigerator to improve noise performance and lower latency for feedback and feedforward experiments [12,13]. Control electronics sends the signals to the millikelvin stage using a small number of lines, where multiplexers and demultiplexers distribute them to many individual quantum devices for increasing characterisation throughput, or to address a single multiqubit chip for quantum information processing applications.

Several approaches to signal multiplexing at millikelvin temperatures have already been explored, however, they do not satisfy the necessary requirements needed for scalable quantum device characterisation or multiqubit control, related to frequency range, insertion loss, isolation and heat dissipation. Commonly used devices for multiplexing signals over a large frequency range are electro-mechanical switches [21–23]. Despite high microwave isolation between ports and no heat dissipation in the idle state, the large physical size and strong Joule heating during state change does not make it suitable for scalable device characterisation [22]. Routing microwave signals at millikelvin temperatures without heat dissipation was demonstrated with nanowires [24] and Josephson-junction devices [25–27], however, these devices have limited bandwidth ($< 2$ GHz) [25,26] and low dynamic range [24]. Very low heat dissipation during switching and no dissipation in idle state was achieved with a phase change switch [28]. Possible creation of infrared radiation due to high local peak temperatures of 1350 K and high insertion loss make phase change switch less suitable for quantum device characterisation. FET transistors based on CMOS [29] or HEMT [30] technology were also used for microwave signal multiplexing, but their implementation exhibited limited bandwidth ($\sim 20$ MHz) [29] and high ($> 10$ dB) insertion loss in the 4-8 GHz frequency range [30]. Scalable DC characterisation was recently demonstrated with CMOS technology. DC multiplexers were either cointegrated with Device Under Test (DUT) [29,31,32], built as a separate unit with off-the-shelf CMOS components [20] or with custom designed CMOS chips [19]. Scalable qubit control electronics was recently demonstrated for spin qubits [33–37], however, high-frequency and high-performance signal multiplexers required for microwave device characterisation and qubit control for superconducting qubit technology have not yet been demonstrated.

Here, we report on a custom designed proof-of-principle cryo-CMOS multiplexer for scalable quantum device characterisation. The multiplexer can be controlled with a parallel or serial interface, has a large bandwidth (DC - 10 GHz), very low insertion loss (1.6 dB at 6 GHz), high isolation (34 dB at 6 GHz), nanosecond switching times and moderate heat dissipation (36.2 µW) at millikelvin temperatures. We demonstrate the use of the multiplexer by characterising intrinsic microwave loss in four superconducting lumped-element resonators (LER) with intrinsic Q-factors ranging from $10^4$ to $7 \cdot 10^6$ at microwave powers corresponding to a single photon in a resonator. The demonstrated scheme simulates short-loop scalable characterisation of microwave loss sources introduced by different fabrication steps. This is critical for fabrication process development of high-performance superconducting qubits. We verify that the residual heat dissipation does not influence measured resonator performance. The demonstrated technology paves the way to scalable quantum device characterisation of resonators [38], microwave kinetic inductance detectors [39], quantum-limited amplifiers [40], as well as superconducting [41] and spin qubits [42]. This work also explores the interaction between cryo-CMOS electronics and superconducting qubits with a perspective of cryo-CMOS-controlled time-division multiplexing of multiqubit control signals.

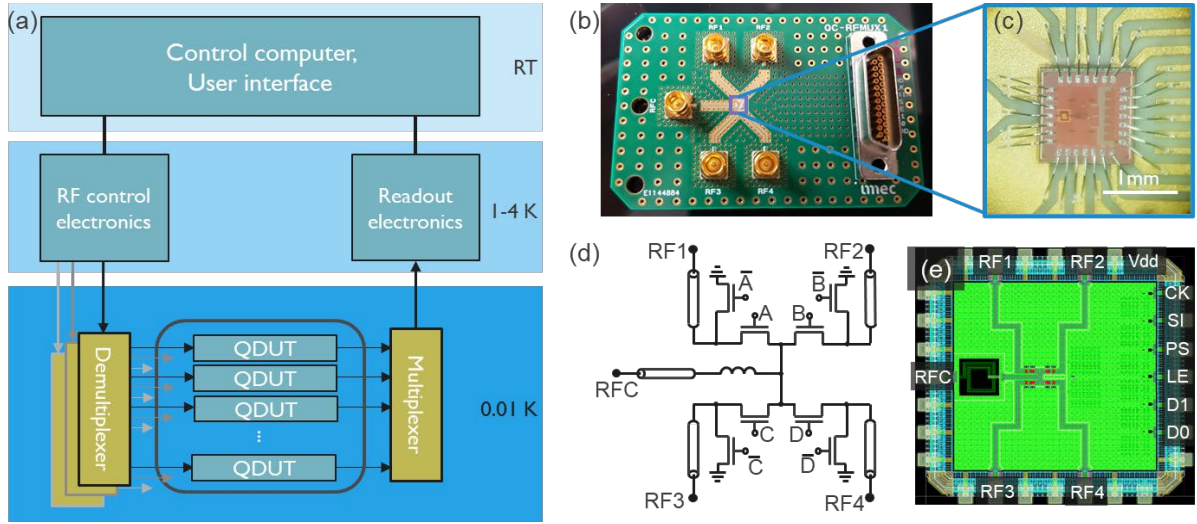

**Figure 1**: Cryo-CMOS multiplexer for microwave signal routing. (a) Schematic of a large-scale characterisation platform for Quantum Devices Under Test (QDUT) or a multiqubit processing platform. Cryo-CMOS multiplexer and demultiplexer reduce the number of required input and output RF lines from the room temperature (RT) to the base temperature stage (0.01K) of a dilution refrigerator. (b) Photograph of a PCB containing the CMOS-multiplexer device. (c) Micrograph of the cryo-CMOS multiplexer chip. (d) Simplified circuit schematic of the multiplexer. RF1-RF4 are the 4 input RF ports, RFC is the common RF output port, A-D are DC or low frequency select lines for the multiplexer. (e) Chip layout showing spatial configuration of the CMOS multiplexer device. Protective input-output (IO) ring (blue, cyan) surrounds the digital control logic core and switch transistors. RF ports and CMOS control lines are denoted at the edge: D0, D1, LE low frequency control lines are used for the parallel interface and CLK, SI, PS, LE for the serial control interface. All ports without a label are connected to $V_{ss}$ (ground).

## 2. Cryo-CMOS multiplexer

A single-pole-4-throw (SP4T) reflective CMOS multiplexer is custom-designed using standard TSMC 28 nm HPC+ technology and optimized for low microwave insertion loss and high isolation between input and output ports (figure 1). The multiplexer device is designed to operate over a wide frequency range, from DC to 10 GHz (figure A1 in supplementary material A), and a wide temperature range. All four input RF ports can be selectively connected or disconnected from the common RF port (RFC) using four pairs of series and shunt switch NMOS transistors, as shown in the simplified circuit schematics in figure 1(d).

A series-shunt switch topology [43] is chosen to increase the isolation despite a small increase in insertion loss caused by the shunt switch transistor. The optimal insertion loss and isolation is realized with minimum length and optimized width of NMOS transistors. The optimal width is chosen such that the "on"-state resistance ($R_{on}$ < 10 Ω) of the switch is small compared to the 50 Ω transmission line impedance, while keeping an acceptably low "off"-state capacitance ($C_{off}$ < 50 fF). To further reduce the insertion loss of the device, a 450 pH inductor is added to the common RF port (RFC) to match the output impedance of the on-chip switches to the 50 Ω impedance of the transmission line (figure 1(d)). Since signal absorption at un-selected ports is not needed for presented application, a reflective switch topology is chosen to maintain low insertion loss in the signal path. The control logic core and large switching transistors are protected by a surrounding input-output (IO) ring, shown in blue and cyan in figure 1(e), which provides protection against electrostatic discharge (ESD). The digital logic and the surrounding IO ring are designed with the standard (room-temperature) libraries. No design optimization is made for cryogenic operations.

The multiplexer can be controlled by either a parallel or a serial interface. The parallel interface requires $\log_2(N)+1$ digital interface lines, where $N$ is the number of multiplexer ports, and a serial interface requires only 4 digital interface lines independent of the number of multiplexer ports. However, the serial interface programming time scales linearly with the number of ports ($t = N \cdot t_{clk}$, where $t_{clk}$ is the clock period), which can become noticeable for large number of ports. The number of digital interface lines in the multiplexer is significantly smaller than the number of lines ($N + 1$) required to operate a mechanical switch, especially for large $N$ [21,22]. In addition to reducing the number of RF lines between room and base temperature plates in a dilution refrigerator, parallel and serial interfaces therefore also drastically reduce the number of digital interface lines. This makes cryo-CMOS multiplexers ideal for large-scale device characterisation or for distributing signals to a large quantum computing processor.

We highlight that the small physical chip size (1.12 x 1.12 mm$^2$) and compact design of presented cryo-CMOS multiplexer (figure 1) allows to allocate more available refrigerator space to Quantum Device Under Test (QDUT), thereby further increasing characterisation throughput. The size of the printed circuit board (PCB), currently set by the size of five SMP connectors, can be further reduced in future designs using multi-port RF connectors or adding QDUTs on the same PCB.

## 3. Cryo-CMOS multiplexer characterisation

The presented cryo-CMOS multiplexer operates over a wide temperature range between room temperature and 32 mK. The lowest temperature is the base temperature of the dilution refrigerator with fully powered multiplexer attached to the base plate. We have thermally cycled multiplexers between room temperature and millikelvin temperatures more than 10 times and did not observe any performance degradation in terms of insertion loss or isolation.

Microwave properties of a multiplexer device are characterised at both room and 32 mK temperature. Characterisation at cryogenic temperature is limited to the 4 – 8 GHz bandwidth by a bandpass filter and a HEMT amplifier mounted on the output line in the dilution refrigerator (figure A2(a) in supplementary material A). The insertion loss is measured as a $S_{21}$ scattering parameter through the multiplexer input port RF4 and output port RFC with RF4 port internally connected to RFC ("on" state). Isolation is measured between the same ports, but with all input ports internally shunted ("off" state).

Room-temperature and low-temperature insertion loss measurements agree well with the room-temperature and -40°C insertion loss simulations as shown in figure 2(a). We note that -40°C is the lowest temperature at which available transistor models are specified. 1-2 dB discrepancy between room-temperature measurements and simulations come from losses in PCB and packaging which are not included in simulations. The insertion loss at 32 mK ranges between 1 and 3 dB with ~1.6 dB at 6 GHz. Similarly, both room- and low-temperature measurements of the isolation agree well with room-temperature and -40°C simulations. The low-temperature isolation ranges between 30 and 40 dB with

~34 dB at 6 GHz (figure 2(c)). Comparable microwave properties confirm that the device performs nominally over a wide operation temperature range down to millikelvin temperatures. In fact, the insertion loss reduces by ~ 0.9 dB at 6 GHz between room and 32 mK temperature. This is attributed to higher conductance of the copper transmission lines on the chip and PCB at low temperatures. The insertion loss is relatively constant in the 4 – 8 GHz frequency range with reduced performance expected above 10 GHz due to the SMP PCB mount connector and bond wires on the PCB (figure A1 in supplementary material A). Isolation shows a noticeable frequency dependence resulting in ~ 10 dB reduction between 4 and 8 GHz, which is within the experimental uncertainty independent of temperature. This behaviour is attributed to the frequency dependent signal leakage through the parasitic capacitance of the series switch transistors in the "off" state.

The multiplexer is designed to operate at a supply voltage of $V_{dd}$ = 0.9 V, however, it is important to study device performance also at reduced $V_{dd}$ in order to explore device limits and power dissipation. When lowering the supply voltage, both insertion loss and isolation degrade (figures 2(b), (d)). At millikelvin temperature, the multiplexer stops operating at $V_{dd}$(stop) = 0.475 V at which point both isolation and insertion loss become 21.4 dB. At $V_{dd}$(stop) the gate-source voltage of NMOS switching transistors falls below their threshold voltage. At room temperature microwave properties degrade slower with lowering supply voltage and reach a lower $V_{dd}$(stop) of ~0.1 V. Room-temperature simulations qualitatively agree with the measurements. Shallower measured sub-threshold slope likely results from a finite distribution of transistors' threshold voltages on a physical device, which is not captured by simulations. The increase of $V_{dd}$(stop) and stronger $V_{dd}$ dependent insertion loss and isolation at lower temperatures is in agreement with the increase of the threshold voltage and steeper subthreshold slope of NMOS transistors at lower temperatures [44,45]. This is also indicated by simulations performed at -40°C.

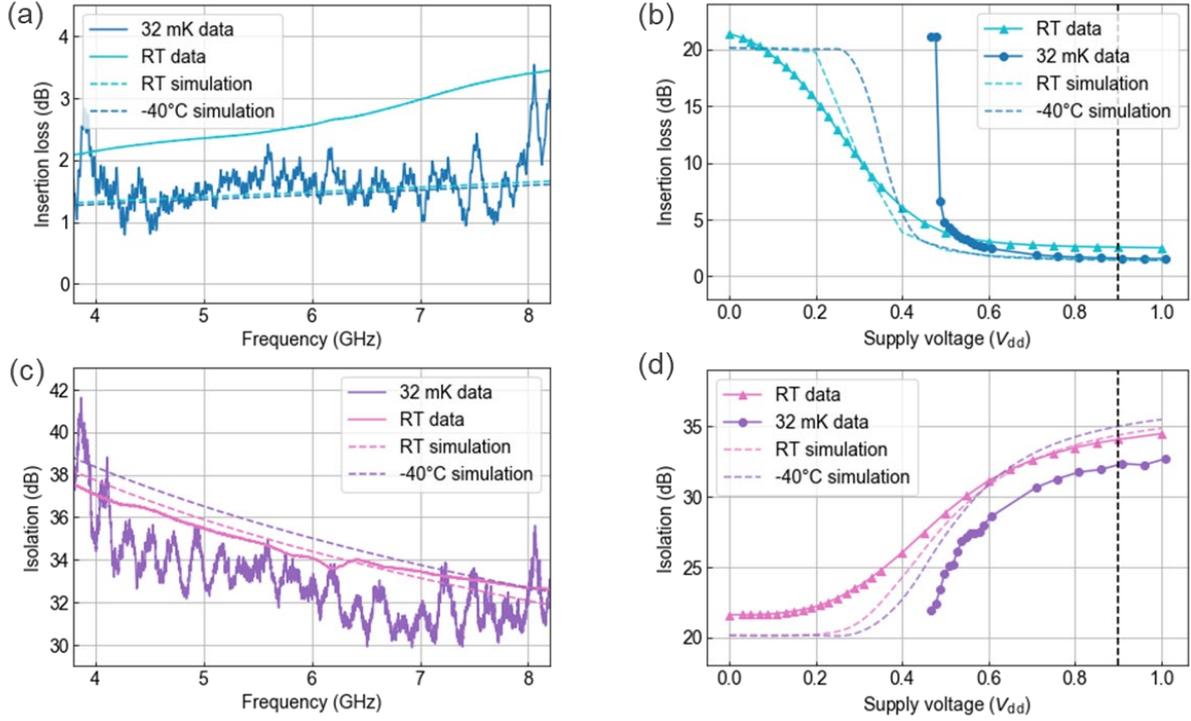

**Figure 2**: RF characterisation of a cryo-CMOS multiplexer. (a) Insertion loss as a function of frequency and (b) supply voltage ($V_{dd}$) at room temperature and base temperature of 32 mK. (c) Isolation as a function of frequency and (d) supply voltage at room temperature and 32 mK. Measurements as a function of supply voltage are performed at 6 GHz. All panels contain simulation results performed for room and -40°C temperature, denoted by dashed lines. Vertical black dashed line in (b) and (d) indicate the operating $V_{dd}$ used for measurements shown in panel (a), (c)**.** Room temperature measurements are performed outside the dilution refrigerator. Notable measurement uncertainty at cryogenic temperatures is due to the unavailable cryogenic calibration standard (see section A in supplementary material).

At 32 mK, the cryo-CMOS multiplexer dissipates 36.2 μW at the nominal supply voltage of $V_{dd}$ = 0.9 V (figure A3(b) in supplementary material A). This is responsible for a slightly elevated dilution refrigerator temperature of 32 mK compared to ~10 mK when the multiplexer is powered off (figure A3(b) in supplementary material A). This static heat dissipation does not depend on the multiplexer state and no dynamic dissipation above the measurement uncertainty of ~15 nW can be observed during operational state switching when using either parallel or series control interface. The operational state switching frequency does not exceed 10 mHz for the demonstrated application. We attribute the residual static dissipation to a small leakage in 126 large clamping transistors in the protective IO ring (see figure 1(e)). To verify this claim, we independently measured current though two parallel ESD protection cells at 4 K. The results show that ESD cell leakage directly scales with $I_{dd}$ measured though the multiplexer at low temperature, confirming our claim (see figure A3(a) in supplementary material A). We note that device simulations with foundry models at -40°C yield approximately two orders of magnitude lower leakage current, in contrast to the experiment. The development of accurate CMOS transistor models at deep cryogenic temperatures is therefore needed and necessary to design and optimize future cryogenic CMOS devices.

To characterise the switching dynamics at millikelvin temperatures, we measured the switching time of the cryo-CMOS multiplexer. The signal rise time is 1.2 ns at 6 GHz and 1.8 ns at 4 GHz (figure A4 in supplementary material A). This rise time is mainly caused by the RC time constant of the NMOS switch transistor. With an optimized biasing circuit, the rise time can be further reduced without increasing the insertion loss in future designs.

## 4. Scalable millikelvin device characterisation

Reducing intrinsic microwave loss in superconducting quantum devices is essential for fabricating high-performing qubits with long coherence time [9]. To study microwave loss in materials, surfaces and interfaces and provide valuable experimental feedback for the fabrication process optimization, many samples must be characterised in a short amount of time. Here we use the cryo-CMOS multiplexer to demonstrate a proof-of-concept multidevice characterisation with a single RF input and a single RF output line in the dilution refrigerator. The characterization throughput is, therefore, increased compared to the scheme where each sample requires its own input and output line in a dilution refrigerator.

In the presented scalable cryogenic device characterisation scheme (figure 1(a) and figure A2(b) in supplementary material A) the cryo-CMOS multiplexer connected to the input line is used for signal demultiplexing and multiplexer connected to the output line is used for signal multiplexing. In this configuration, we study microwave loss using four samples containing superconducting high-quality factor LER fabricated with different process steps (table 1). Resonator are widely used as short-loop characterization vehicles for superconducting qubit devices, since their lifetime time is limited by the same dominant microwave losses as in qubits [46]. Qubit specific loss sources, such as quasiparticle tunnelling or TLS defects inside a Josephson junction, are not captured by this method, however these are generally not limiting the performance of widely used Transmon qubits [47]. A short-loop study can provide fast and reliable feedback necessary for rapid fabrication process development.

The intrinsic microwave loss is characterised by the internal quality factor $Q_i$ of the resonator [9]. Samples are inserted and measured in high-purity superconducting aluminium 3D cavities, which enable fast sample exchange (see figure 3(a) and supplementary material A). Due to wireless coupling between LERs and the electric field of the 3D cavity's fundamental mode, samples do not need to be glued or wire bonded. The superconducting 3D cavity ensures excellent shielding against external static and dynamic magnetic fields. It also provides a well-defined electromagnetic environment void of unwanted parasitic modes that typically lead to additional microwave losses in coplanar superconducting qubits and resonators [48,49].

Each sample contains 6 LERs composed of a meander inductor and a dipole-antenna-shaped capacitor (figure 3(b)). The geometry of a capacitor and the LER position within the cavity determines the coupling between the LERs and the fundamental mode of the 3D cavity [50]. For the demonstration of proof-of-concept signal multiplexing, we present only a single LER at 4.8 GHz on each sample. We also choose four fabrication processes that yield samples with large spread in Q-factor values to test if cryo-CMOS multiplexers can be used for characterizing resonators in the full range of their performance parameters.

**Table 1**: List of samples being measured with the CMOS multiplexers. Frequencies and measured low power Q-factors of resonators shown in figure 3 are listed in the right two columns.

| Sample name | Fabrication details | Resonator frequency | Low power Q-factor |
|---|---|---|---|
| S1 | Nb on standard Si substrate (1-100 Ω cm) | 4.802 GHz | $(12 \pm 2) \cdot 10^3$ |
| S2 | Nb on high resistivity substrate (> 3.3 kΩ cm). Structures were in-situ coated with 20 nm of $SiN_x$ in a deposition chamber | 4.815 GHz | $(200 \pm 50) \cdot 10^3$ |
| S3 | Nb on high resistivity substrate (> 3.3 kΩ cm) | 4.803 GHz | $(1.5 \pm 0.5) \cdot 10^6$ |
| S4 | Nb on high resistivity substrate (> 3.3 kΩ cm). Sample is cleaned with 5% wt. HF solution for 60 s, loaded in the fridge and cooled down in less than 20 min. | 4.779 GHz | $(7 \pm 2) \cdot 10^6$ |

Transmission spectrum measurements through cryo-CMOS multiplexer and demultiplexer show that LERs exhibit nearly ideal Lorentzian line shapes at millikelvin temperatures (figure 3(d)). The resonances are slightly asymmetric due to the coupling to the 3D cavity mode offset by more than 2 GHz towards higher frequencies. Nevertheless, the transmission spectra can be well fitted using generalized Lorentzian line shape with a complex coupling quality factor (see supplementary material A). Setting the coupling quality factor $Q_c$ to be much larger than the highest expected internal $Q_i$ factor allows us to measure internal microwave losses ($1/Q_i$) originating from defects in the substrate and different interfaces on a device (figure 3(c)) by directly measuring loaded quality factor $Q_L$ (figure 3(d)), due to $1/Q_L = 1/Q_c + 1/Q_i \approx 1/Q_i$.

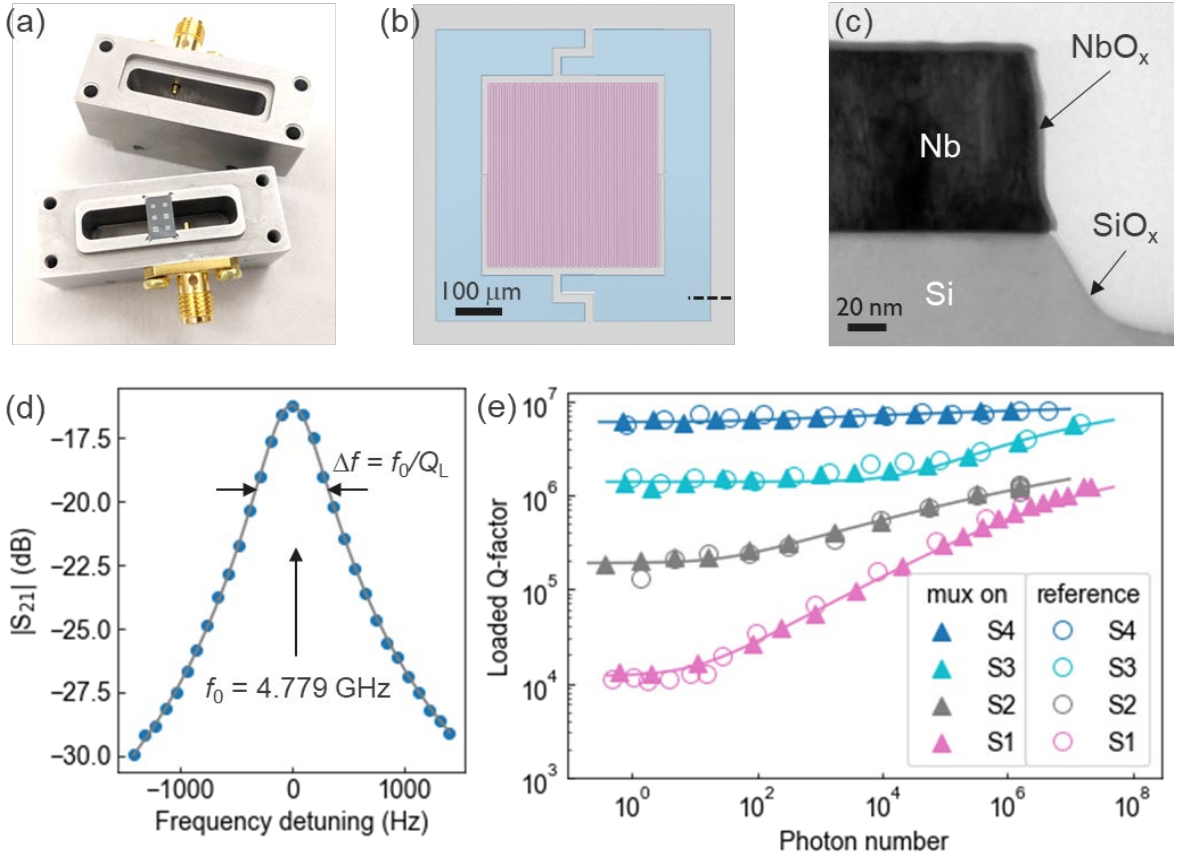

**Figure 3**: Lumped element resonator (LER) device characterisation. (a) Photograph of an aluminium 3D cavity containing a sample with 6 LERs. (b) False-colour micrograph of a LER. Meander inductor and antenna-shaped capacitor are depicted in red and blue, respectively. Silicon substrate is shown in grey. (c) Cross-TEM micrograph of the metal-substrate-air interface. The cross-section location is indicated with a dashed line in panel (b). TLS defects are normally present at the interfaces, metal-air ($NbO_x$), substrate-air ($SiO_x$), substrate-metal or in the Si substrate [9]. (d) Magnitude of a transmission spectrum for the selected LER in sample S4 measured using CMOS demultiplexer - multiplexer configuration at high photon number powers. Grey solid line shows a fit described in supplementary material A. (e) Loaded quality factor of four samples measured using CMOS multiplexers as a function of drive power expressed in number of photons in the resonator. Data was fitted with equation (A1) in supplementary material A. Reference measurement results, obtained with unpowered CMOS multiplexers, are denoted with empty symbols.

Loaded quality factors are extracted by fitting power-dependent transmission spectra of the four samples (see supplementary material A). For convenience, we express the applied microwave power in terms of the number of microwave photons in a LER (see supplementary material A and C). We measure the power dependence of $Q_L$ down to a single-photon level power in the regime, where superconducting qubits and other quantum devices operate (figure 3(e)). At millikelvin temperatures and single-photon level powers, the dominant microwave loss is attributed to two-level-system material defects found predominantly in amorphous interfaces layers [9]. These exhibit a power-dependent loss due to the two-level-system (TLS) nature of the defects. The combined loss is described by the following expression [9,51]:

$$\frac{1}{Q_i} = \sum_i \frac{p_i \tan\delta_i}{\left(1 + \frac{n}{n_{ci}}\right)^{\beta_i}} + \frac{1}{Q_0}. \tag{1}$$

Here $p_i$ is the participation ratio of the $i$-th loss component, which is equal to the amount of electric field energy stored in that region relative to the total electric-field energy of the system. $\tan\delta_i$ is the loss tangent of the $i$-th component, $n$ is a number of photons in the resonators, $n_{ci}$ is a critical number of photons in resonators above which the TLSs start to saturate, $\beta_i$ is an exponent dependent on the electric field distribution along the considered region [9] as well as the TLS-TLS interaction ($\beta = 0.5$ when interaction is absent) [9,52], and $Q_0$ is power independent loss that can include residual dielectric, quasiparticle or radiation loss [9].

The four samples exhibit notably different power or photon number dependent quality factors, nevertheless, they can be well fitted with Eq. (1) using a single effective TLS component. The highest quality factor of $(7 \pm 2)\cdot 10^6$ at single photon level with the weakest power dependence is found for sample S4, where the native silicon and Nb oxides are removed by submerging the sample in hydrofluoric (HF) acid approximately 20 min before the cooldown. Remaining power dependent losses are likely attributed to the residual TLS losses in substrate-metal interface or high-resistivity silicon substrate [38,53]. A lower, but still high-quality factor of $(1.5 \pm 0.5)\cdot 10^6$ is found in S3, where a combination of native surface oxides and the substrate limit the loss. To prevent native oxidation of surfaces exposed to air, a 20 nm layer of SiN$_x$ was in-situ coated on top of the sample S2 immediately after etching. Despite the absence of metal oxides, the measured low-power quality factor of (200 ± 50)·10$^3$ is lower than that of S3. Using Eq. (1), the simulated participation ratio of the SiN$_x$ layer ($p_{SiNx}$ = 0.21%) and neglecting losses from the substrate and metal-substrate interfaces, we obtain a loss tangent $\tan\delta_{SiN_x} \approx 2.3·10^{-3}$. The extracted loss tangent of SiN$_x$ is higher than what was published previously [54], likely due to the lower nitride deposition temperature used for our samples. The strongest power dependence and the lowest single photon quality factors of (12 ± 2)·10$^3$ are found in S1. Here, a standard silicon substrate with >1 Ω·cm resistance was used. Using a participation ratio of $p_{Si}$ = 0.917 for the silicon substrate, we can extract the loss tangent of standard silicon wafer to be $\tan\delta_{Si} = (9 \pm 2) \cdot 10^{-5}$. Participation ratio simulation results and calculation details can be found in supplementary material B.

The residual power dissipation of CMOS multiplexers can increase the electron temperature of a sample and therefore increase static thermal photon population in LER. To verify whether the CMOS device dissipation affects resonator characterisation, we compare the presented results to reference quality factors measured through the multiplexer and the demultiplexer in an unpowered state ($V_{dd}$ = 0). We note that the measurement uncertainty is larger especially at low photon numbers, due to an order of magnitude smaller signal to noise ratio resulting from 21.4 dB insertion loss of the unpowered multiplexer. Within the experimental uncertainty, the reference Q-factors follow the same power dependence as those measured through multiplexers (empty markers in figure 3(e)), even for S1 where the power dependence is largest and quality factors show a clear saturation at the single photon level powers. Since no discrepancy between measured and reference $Q_L$ can be observed down to the single photon level, we can surmise that residual thermal occupation in the resonator must be less than ~1 photon.

To measure the effective temperature of a resonator connected to a cryo-CMOS multiplexer, we measure an ac Stark shift of a Transmon qubit coupled to a coplanar waveguide resonator [55]. At the operating supply voltage of $V_{dd}$ = 0.9 V, the ac Stark shift of a qubit corresponds to a mean photon number of 2.2 photons in the resonator (see supplementary material D). This is in agreement with the power dependent quality factor measurement presented above. The average photon number in a resonator corresponds to an electron temperature of 0.83 K. We can conclude that the residual heat dissipated by the CMOS multiplexers does not influence the high-Q factor resonator characterisation.

## 5. Discussion

The presented cryo-CMOS multiplexer is a proof-of-concept device designed for scalable characterisation. The SP4T multiplexer has a modular circuit design and can be expanded to any number of input RF ports. In the current implementation the assembled multiplexer physical size is set by the RF PCB mount connectors, as can be seen in figure 1(b). When scaling to a large number of multiplexer ports, static and dynamic heat dissipation will remain unchanged, since both number of ESD cells and simultaneous transistor switching do not scale with the number of ports in this application. RF crosstalk between closely spaced RF pins on a large-scale cryo-CMOS chip could increase, however, increasing the chip size or using vertical interconnect such as flip-chip technology is commonly used to alleviate this problem. When considering the entire scalable cryogenic device characterization setup, most of the space would be taken by individual device sample holders (figure 3(a)) needed for shielding against environmental noise. To further increase the measurement throughput, quantum devices could be mounted on a specially designed large samples holders with a common PCB containing cryo-CMOS multiplexers and several quantum devices while providing required shielding. Taking advantage of multilayer PCB wiring, daughter PCBs [19], or high-density flex line connections for dense signal routing and thermal management [15] would further aid the scaling process.

Current multiplexer implementation is not yet compatible with scalable high-fidelity qubit control or complete qubit characterisation, due to noticeable thermal radiation emanating from the multiplexer. However, based on the presented heat dissipation results, we can identify two possible directions that can substantially reduce thermal radiation of the cryo-CMOS multiplexer and with that open a path to scalable high-fidelity qubit control.

The first direction is to reduce the number of leaky ESD protection units in the design from currently 126 to the minimum safe value of 4. Assuming that majority of dissipation originates from the protection ESD circuits, reducing the number of ESD protection units by a factor of 31.5 would lower the heat dissipation from 36 μW to 1.1 μW. The estimated thermal population in the resonator would then be lowered from ~2 photon to ~0.06 photons, which is comparable to the residual thermal population in the state-of-the-art qubit devices [56]. This is a scalable solution since the residual thermal dissipation does not dependent on the number of multiplexer ports.

The second direction is to thermalize the signal at the output of the multiplexer by a well thermalized attenuator as it is typically done for signal input lines in a dilution refrigerator [14]. Due to a sufficient dynamic range of CMOS multiplexers the input signal power can be increased to compensate for the signal attenuation at the output. The attenuation can be either done by the off-the shelf or custom designed cryo-attenuators [57] or by PCB-mounted attenuators. In the latter case thermal isolation between the "hot" CMOS part of the PCB and "cold" qubit part of the PCB can be achieved with superconducting high-density rigid-flex throughs [15].

A combination of the proposed two directions can lead to cascaded thermal radiation reduction by more than 4 orders of magnitude at the expense of a redesign of the circuit or additional microwave components. To go beyond this restriction and achieve even lower dissipation, other techniques are also worth exploring such as cryo-CMOS with optimized switching voltages [33], adiabatic switching [58], or using technologies such as FD-SOI CMOS [59] or superconducting SFQ electronics [60].

# 6. Conclusions

We present custom designed cryo-CMOS multiplexers with very low insertion loss and high isolation over a wide frequency range that are well suited for scalable characterisation of devices, such as superconducting resonators, quantum-limited amplifiers, microwave kinetic inductance detectors, and to an extend superconducting and spin qubit devices.

We emphasize that the presented cryo-CMOS multiplexers are designed and fabricated using a commercially available CMOS technology, and do not require additional technology development. Based on demonstrated nanosecond switching capabilities, high isolation, low insertion loss, and possibility to further reduce heat dissipation, we can conclude that cryo-CMOS multiplexers represent a viable solution for increasing the cryogenic characterisation throughput and can contribute to alleviating the wiring problem in future large-scale quantum computing systems.

# Acknowledgments

The authors gratefully thank the imec P-line, operational support, and the MCA team. This work was supported in part by the imec Industrial Affiliation Program on Quantum Computing. The authors would also like to thank prof. A. Wallraff for providing lumped element resonator designs.

## A. Experimental setup, device simulation and characterisation

Experiments at cryogenic temperatures are performed in a Bluefors LD400 dilution refrigerator where input lines are attenuated at various temperature stages and connected to a demultiplexer device (see figure A2). The output port of the multiplexer is connected to the fridge output line containing three LNF-ISC4_8A isolators and a KBF-4/8-2S 4-8 GHz bandpass filter at the base and cold plate stage. The signal is first amplified at 3 K using an LNF-LNC4_8C HEMT amplifier and at room temperature using a Miteq LNA-30-04000800-07-10P ultra low noise amplifier (see figure A2).

Samples containing lumped-element resonators are placed in an aluminium 3D cavity at the electric field maximum of the cavity's TE101 mode. To achieve low intrinsic loss of 3D cavities, high purity aluminium is used (AW-1350A), and the cavity surfaces are cleaned with Honeywell phosphoric acid etching mixture for 15 min. When closing the cavity, an indium wire (99.99%) seal is used to ensure light-tight enclosure and a superconducting electrical contact. The length of the two SMA connector pins reaching inside the 3D cavity and LER coupling to the TE101 mode are optimized to yield LER coupling quality factor of $\sim 50 \cdot 10^6$, which is much higher than expected internal quality factors of LERs.

The multiplexer chip is placed on a four-layer printed circuit board (PCB) with low loss Rogers RO4350B microwave laminate. Two identical multiplexer PCBs, one functioning as demultiplexer, are mounted on the mixing-chamber plate of a dilution refrigerator and thermalized with oxygen-free copper braids. Superconducting devices under test are placed in separate sample holders (3D cavities) and are connected to RF ports of the multiplexer with semirigid copper coaxial cables and flexible cryogenic RF cables (Cri/oFlex® 2).

The CMOS multiplexer was controlled with a programmable voltage source BiLT iTest BE2142. All ports shared the same floating ground that was connected to the fridge ground potential.

High Q-factor LER and CMOS multiplexers are characterised using a Keysight P5004 vector network analyser. The coaxial lines were calibrated using a Keysight ECal N4431 for room temperature scattering parameter measurements. At millikelvin temperatures the insertion loss of the device is measured as a difference between S21 scattering parameter measured in dB with a device and without a device measured independently. In the latter case, input and output cables are connected with a single SMP-SMP female-female bullet (figure A2). We note that insertion loss extracted in this way constitutes of a loss in the multiplexer, PCB lines and SMP connectors. The extracted insertion loss of the multiplexer is expected to be lower than the reported value. Isolation is determined as an insertion loss through a device in the "off" state. We added approx. sign (~) to the reported insertion loss and isolation values to indicate that uncertainty is not well defined.

LER spectra (shown in figure 3(d)) are fitted using Lorentzian line shape with a complex external quality factor [1]:

$$S_{21}(f) = A \frac{Q_L/|Q_c|e^{i\varphi}}{1 + 2iQ_L\left(\frac{f}{f_r} - 1\right)} + B, \tag{A1}$$

where $A$ and $B$ are complex constants, $f_r$ is the LER's resonance frequency, $Q_L$ is the loaded quality factor, $Q_c$ is the complex coupling quality factor and $\varphi$ is the phase of $Q_c$, associated with line shape asymmetry. $Q_L$ is a function of a coupling quality factor and the internal quality factor, $1/Q_L = 1/|Q_c| + 1/Q_i$. The coupling quality factor of a LER is determined by the Purcell decay [2] though the 3D cavity. We have engineered the coupling quality factor to be much larger than expected internal quality factors ($Q_c \sim 50 \cdot 10^6$). In this setting $Q_L$ is equal to the internal quality factor $Q_i$. This is the reason why in the main text we use $Q_L$ to measure $Q_i$.

The number of microwave photons in a LER is calculated using equation (C12) (see supplementary material C). The microwave power at the chip was calculated based on the output power of the instrument and the attenuation in the input lines (figure A2).

Cryo-CMOS Multiplexer design was optimized for low insertion loss and high isolation by performing room-temperature and -40°C device simulations. -40°C is the lowest temperature at which available transistor models are reliable. Chip level circuit simulations are done in Cadence Virtuoso. Electromagnetic simulations are performed using Integrand EMX for the on-chip passives.

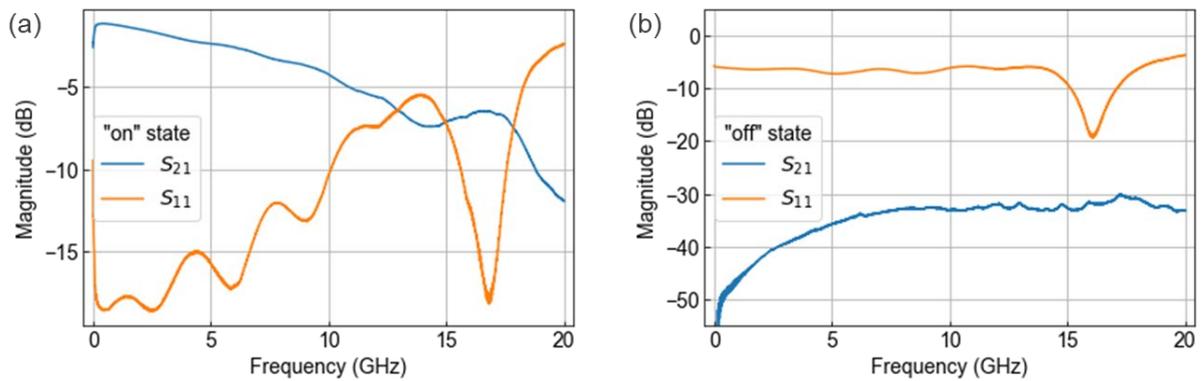

**Figure A1**: Room temperature S-parameter characterisation of the cryo-CMOS multiplexer though RF4 and RFC ports. (a) Insertion loss ($S_{21}$) and reflection ($S_{11}$) when the multiplexer is in the "on" state. See main text for details. (b) isolation ($S_{21}$) and reflection ($S_{11}$) for multiplexer in the "off" state. Both measurements are performed at RT outside the dilution refrigerator with the supply voltage $V_{dd} = 0.9$ V applied to the multiplexer.

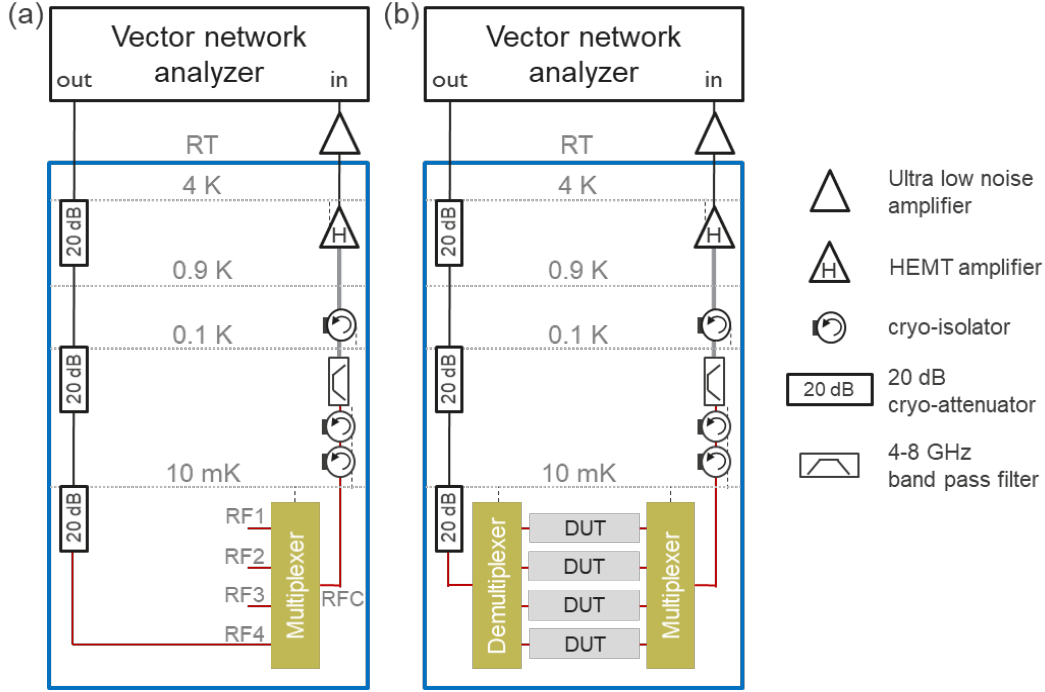

**Figure A2**: Schematics of the experimental setup. DC control lines for multiplexer device(s) are omitted for clarity. (a) Setup used to characterise microwave properties of a single cryo-CMOS multiplexer. (b) Setup used to demonstrate characterisation of high-Q factor resonator using a cryo-CMOS multiplexer and demultiplexer. Details of the experimental setup are presented in the supplementary material A.

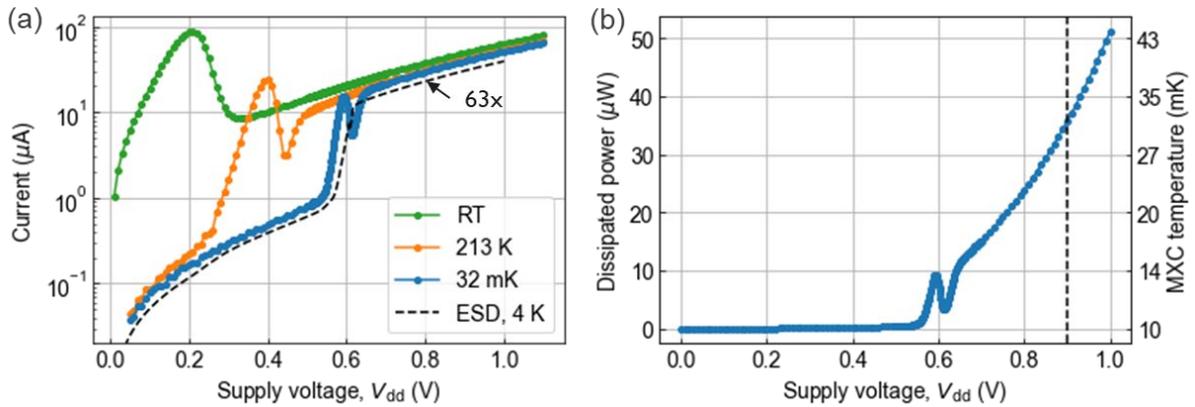

**Figure A3**: Current-voltage measurements and power dissipation of a multiplexer. (a) $V_{dd} \rightarrow V_{ss}$ current as a function of supply voltage $V_{dd}$ at three dilution refrigerator base-plate temperatures. The peak at lower $V_{dd}$ is caused by partial opening of ESD voltage clamping transistors. The increased leakage current visible at higher $V_{dd}$ and lower temperatures is caused by a tunnelling current from drain to gate in the same ESD clamping transistors. Black dashed line corresponds to a current measurement though two individual parallel ESD protection units scaled to the number of ESD units in the device (by a factor of 63) at 4 K. The agreement between the low temperature leakage current though the device and scaled ESD test unit current confirms that leakage originates in ESD protection cells. (b) Dissipated power ($V_{dd} \cdot I$) as a function of supply voltage $V_{dd}$ at millikelvin temperature. Vertical dashed line marks the nominal operating voltage of $V_{dd} = 0.9$ V where heat dissipation is equal to 36.2 µW, as reported in the main text.

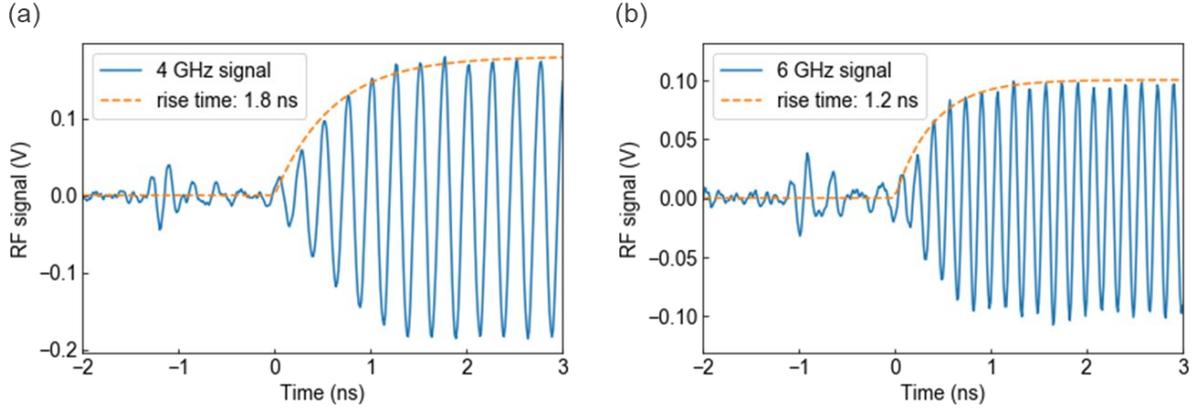

**Figure A4**: CMOS switch rise time at 32 mK. Measurement was performed by sending a coherent tone at (a) 4 GHz and (b) 6 GHz frequency to the input of the CMOS multiplexer and directly detecting the amplified output signal with a real time oscilloscope (Keysight DSOV084A 80GS/s). Orange curve represents an exponential fit to the envelope of the data. The rise time is defined as a time in which the signal reaches 95% of its maximal value. The residual signal pulse cca. 1 ns before the switching event is caused by the voltage rise at the gate of the switching transistors.

### B. Participation ratio simulations

The participation ratio (PR) of dielectric regions is estimated from the electric field distributions obtained from numerical simulations of the design [3]. As opposed to coplanar waveguide resonators (CPW), LER do not possess translational symmetry and cannot be accurately simulated using a 2D cross-sectional simulation. Moreover, a full-scale 3D simulation of a LE resonator presents significant numerical challenges due to the large variation (~ nm to several 100 µm) in the length scales associated with the design. Therefore, we adopt a 2-step approach to reduce the complexity of the problem and to obtain the PRs.

First, a coarse 3D simulation of a LER is performed in the high-frequency electromagnetic solver Ansys HFSS using the eigenmode solution setup. The LER is modelled as a 2D metal sheet on a silicon substrate and is placed in a 3D air box. This enables an accurate simulation of the electric field distribution in the resonator at its resonance frequency as shown in figure B1. The electric field is mainly concentrated around the capacitance region. However, there is also a significant part of the field in the inductor (meander) region. This validates our previous statement that a 2D cross-sectional simulation of the LER is insufficient to accurately estimate the PR. Furthermore, the variation of the electric field over different regions of the design results in different PRs. In order to tackle this problem, several sub-regions are defined in the design as shown in figure B1 and the electric field distribution is used to estimate the fraction of the total energy present in the sub-region.

Next, a fine 2D simulation of each of the sub-regions is performed in Sentaurus TCAD. For each sample (refer to table 1 in the main text), we include the appropriate dielectrics with their estimated thicknesses from physical characterisation [4] in the simulation. Using this, we obtain the refined electric field distribution and, subsequently, the PR for each sub-region. As an example, figure B1 shows the simulation results obtained for sample S1/S3 for the meander-capacitance region.

As a final step, we obtain the total PR of a dielectric as a sum of the PR in various sub-region, weighted with the fraction of the energy distribution obtained from HFSS. The results obtained for the different samples are summarized in table B1.

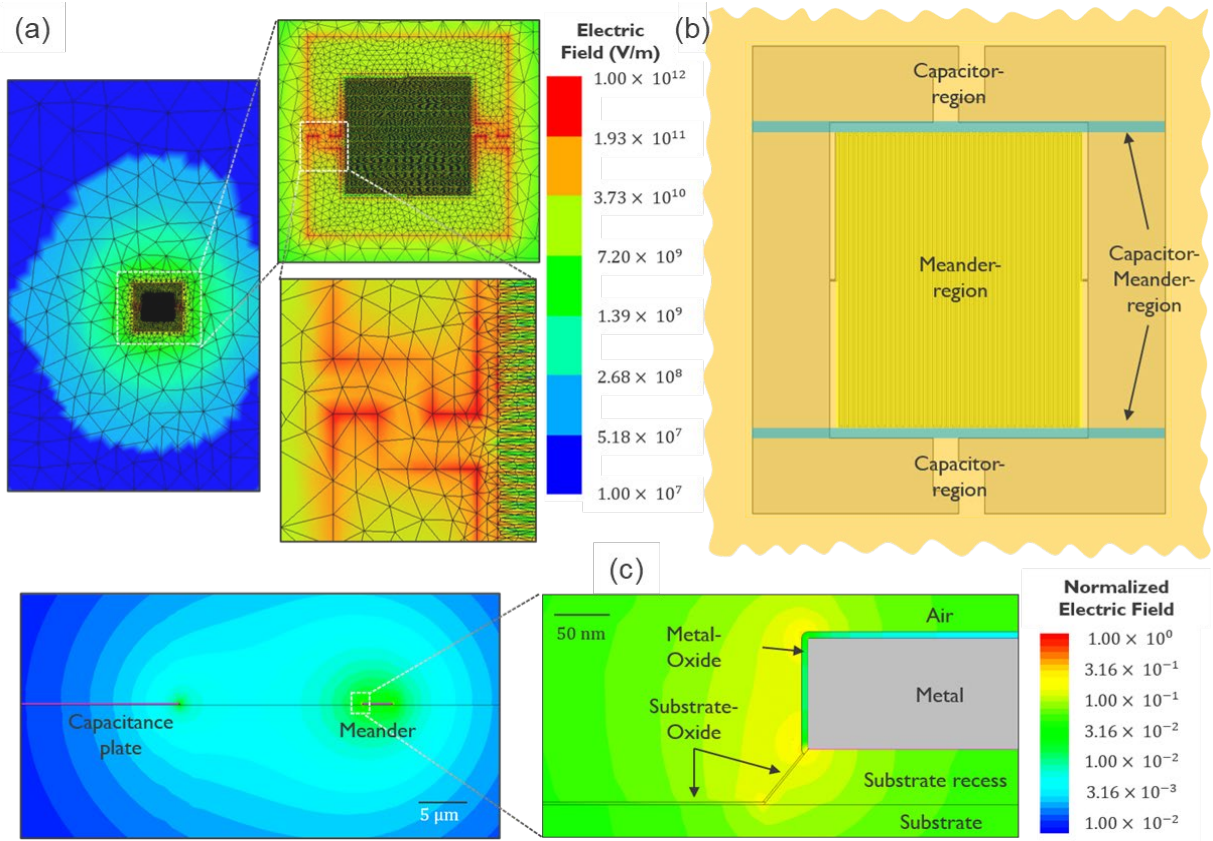

**Figure B1**: Simulation results for estimating the participation ratios. (a) Electric field distribution obtained from eigenmode solution of Ansys HFSS for a LER on a silicon substrate placed in a 3D cavity. Although most of the field is concentrated near the capacitance plates, a substantial amount of field is present in the inductance (meander) region. (b) Regions identified to calculate the fraction of the total energy present. The capacitance (light orange) region has ∼ 78.1% , the meander (light yellow) has ∼16.9% and the capacitance-meander region (light blue) has ∼5.1% of the energy. (c) Electric field distribution (normalized) obtained from Sentaurus TCAD for the capacitor-meander region. The widths of the meander conductor and the capacitor plate is taken as 3 μm and 300 μm, respectively, and have a spacing of 18.5 μm between them. Both conductors have a thickness of 100 nm. The thicknesses of the metal-oxide and the substrate-oxide are 6 nm and 2 nm [4], respectively. Similar simulations have also been performed (not shown here) for the capacitance region (conductor widths 300 μm and spacing 50 μm) and meander region (conductor widths 3 μm and spacing 3 μm). For the simulations, the dielectric constants for silicon substrate ($\varepsilon$ = 11.9), substrate oxide ($\varepsilon$ = 4), metal oxide ($\varepsilon$ = 10) and $SiN_x$ ($\varepsilon$ = 7) are used.

|  | S4 - HF dipped Nb sample | | S3 - Nb sample | | S2 - capped Nb sample | | S1 - standard Si substrate sample | |
|---|---|---|---|---|---|---|---|---|
|  | waPR | loss | waPR | loss | waPR | loss | waPR | loss |
| **Substrate** | 91.4% | $2.38 \cdot 10^{-7}$ | 91.7% | $2.38 \cdot 10^{-7}$ | 91.7% | $2.39 \cdot 10^{-7}$ | 91.7% | $8.34 \cdot 10^{-5}$ |
| **Metal oxide** | 0.00033% | $3.30 \cdot 10^{-9}$ | 0.006% | $5.76 \cdot 10^{-8}$ |  |  | 0.006% | $5.76 \cdot 10^{-8}$ |
| **Substrate oxide** |  |  | 0.012% | $2.02 \cdot 10^{-7}$ |  |  | 0.012% | $2.02 \cdot 10^{-7}$ |
| **Capping layer** |  |  |  |  | 0.21% | $4.77 \cdot 10^{-6}$ |  |  |
| **Total loss** |  | $2.41 \cdot 10^{-7}$ |  | $4.98 \cdot 10^{-7}$ |  | $5.00 \cdot 10^{-6}$ |  | $8.37 \cdot 10^{-5}$ |
| **Q-factor** |  | $4.15 \cdot 10^{6}$ |  | $2.01 \cdot 10^{6}$ |  | $2.00 \cdot 10^{5}$ |  | $1.20 \cdot 10^{3}$ |

**Table B1**: Summary of simulated weighted average participation ratios (waPR) and calculated loss of various dielectric regions on a sample. For each sample total microwave loss and the internal Q-factor are calculated at the bottom. Loss is calculated as a product of the average participation ratio and loss tangent of that region. The following loss tangent were used in the calculations: high-resistivity silicon substrate ($\tan\delta = 2.60 \cdot 10^{-7}$) [5], metal oxide ($\tan\delta = 1 \cdot 10^{-3}$) [6], silicon oxide ($\tan\delta = 1.7 \cdot 10^{-3}$) [5], The loss tangent of SiN$_x$ cap was determined from the measured Q-factor for S2 ($\tan\delta = 2.3 \cdot 10^{-3}$). Loss tangent of standard silicon substrate was determined from the measured low power Q-factors for S1 [$\tan\delta = (9 \pm 2) \cdot 10^{-5}$]. For S4 and S3 estimated Q-factor values agree with measured results presented in figure 3(e) and table 1. We note that loss from the metal-substrate interface is not considered, since it cannot be distinguished from the loss in the substrate [4].

## C. Purcell decay and photon number occupation of lumped element resonators

The dynamics of a lumped element resonators coupled to a 3D cavity can be captured by the Jaynes Cumming Hamiltonian in the rotating wave approximation:

$$\mathcal{H} = \hbar\omega_c a^\dagger a + \hbar\omega_r b^\dagger b + \hbar g \left(a^\dagger b + b^\dagger a\right), \tag{C1}$$

where $\omega_c$ and $a$ are the fundamental mode frequency and field operator of a 3D cavity, respectively, $\omega_r$ and $b$ are frequency and field operator of a lumped element resonator, respectively, and $g$ is the coupling constant.

We write Langevin equations for the two fields ($c = a, b$),

$$\frac{dc}{dt} = -\frac{i}{\hbar}[c, \mathcal{H}] + \text{NU}, \tag{C2}$$

where NU are non-unitary terms describing loss or excitation. Assuming oscillating field operators, $c(t) = c \exp(-i\omega t)$, we arrive at following Langevin equations

$$-i\omega a = -i\omega_c a - igb - \frac{\kappa_i + \kappa_o + \gamma_c}{2} + \sqrt{\kappa_i}\, a, \qquad (C3)$$

$$-i\omega b = -i\omega_r b - iga - \frac{\gamma_r}{2}. \qquad (C4)$$

Here $\kappa_{i,o}$ is the input/output coupling rate between the cavity mode and the coaxial waveguide. These coupling rates can be tuned by adjusting the length of the pin of the SMA panel mount connector inside the 3D cavity. In the experiments we set them to equal values. $\gamma_{c,r}$ is the internal loss rate of the cavity/resonator. Since $\gamma_c \ll \kappa_{i,o}$ we will neglect $\gamma_c$ in final expressions.

Using the two Langevin equations we can express resonator field operators as

$$b = \frac{\dfrac{g\sqrt{\kappa_i}\, a_{in}}{\dfrac{\kappa_i + \kappa_o + \gamma_c}{2} + i(\omega_c - \omega)}}{\omega - \omega_r + \dfrac{i}{2}\left(\gamma_r + \dfrac{2g^2}{\dfrac{\kappa_i + \kappa_o + \gamma_c}{2} + i(\omega_c - \omega)}\right)}. \qquad (C5)$$

Assuming $\omega$ is near $\omega_r$ and resonator-cavity detuning $\Delta = \omega_c - \omega_r \gg \kappa_i, \kappa_o, \gamma_r$, we can simplify the above expression to

$$b = \frac{i\sqrt{\kappa_{Pur\_i}}\, a_{in}}{\omega - \widetilde{\omega}_r + i\dfrac{\kappa_{Pur} + \gamma_r}{2}}. \qquad (C6)$$

Here we define the Purcell decay [7] as $\kappa_{Pur\_i,o} = \kappa_{i,o} g^2/\Delta^2$, $\kappa_{Pur} = \kappa_{Pur\_i} + \kappa_{Pur\_o}$ and $\widetilde{\omega}_r = \omega_r - g^2/\Delta$.

Next, we calculate the spectral shape of a LER inside the 3D cavity. We first express the cavity field as

$$a = \frac{-igb + \sqrt{\kappa_i}\, a_{in}}{\dfrac{\kappa_i + \kappa_o + \gamma_c}{2} + i(\omega_c - \omega)}. \qquad (C7)$$

Inserting equation (C6), assuming $\Delta = \omega_c - \omega_r \gg \kappa_i, \kappa_o, \gamma_r$ and that $\omega$ is near $\omega_r$, we can arrive at

$$a = i\left(\frac{\frac{g\sqrt{\kappa_{\text{Pur\_i}}}}{\Delta}}{\omega - \widetilde{\omega}_r + i\frac{\kappa_{\text{Pur}} + \gamma_r}{2}}\right)a_{\text{in}}. \quad (C8)$$

We proceed by using input-output relation [8] $a_{\text{out}} = \sqrt{\kappa_i}a$ for the output port and express the transmission coefficient as

$$S_{21}(\omega) = \frac{a_{\text{out}}}{a_{\text{in}}} = \frac{i\sqrt{\kappa_{\text{Pur\_i}}\kappa_{\text{Pur\_o}}}}{\omega - \widetilde{\omega}_r + i\frac{\kappa_{\text{Pur}} + \gamma_r}{2}}. \quad (C9)$$

The above expression is equivalent to a resonator coupled directly to a transmission line, with renormalized input and output coupling rates and shifted resonance frequency $\widetilde{\omega}_r = \omega_r - g^2/\Delta$.

The mean number of photons in a cavity can be calculated using Eq. (C6)

$$\langle n \rangle = b^\dagger b = \frac{\kappa_{\text{Pur\_i}}}{(\omega - \widetilde{\omega}_r)^2 + \left(\frac{\kappa_{\text{Pur}} + \gamma_r}{2}\right)^2}|a_{\text{in}}|^2. \quad (C10)$$

We can write incoming photon flux as $|a_{\text{in}}|^2 = P_{\text{in}}/\hbar\widetilde{\omega}_r$ and evaluate the above equation at $\omega = \widetilde{\omega}_r$:

$$\langle n \rangle = \frac{\kappa_{\text{Pur\_i}}}{\left(\frac{\kappa_{\text{Pur}} + \gamma_r}{2}\right)^2}\frac{P_{\text{in}}}{\hbar\widetilde{\omega}_r}. \quad (C11)$$

This can be further simplified assuming $\kappa_{\text{Pur\_i}} = \kappa_{\text{Pur\_o}}$ and expressed with only directly measurable quantitates as:

$$\langle n \rangle = 2\, S_{21}(\widetilde{\omega}_r)Q_L\frac{P_{\text{in}}}{\hbar\widetilde{\omega}_r^2}, \quad (C12)$$

where $Q_L = \widetilde{\omega}_r/(\kappa_{\text{Pur}} + \gamma_r)$ is the loaded-quality factor. equation (C12) is used in the main text to calculate the average photon number in a LER for a specific input power at the sample.

### D. Thermometry with a Transmon qubit

Residual thermal noise irradiated from the cryo-CMOS multiplexer is estimated using a Transmon qubit coupled to a readout resonator. Thermal noise from the cryo-CMOS multiplexer in an "off" state is routed together with control signal to the feedline on a sample containing a coplanar resonator coupled to a Transmon qubit. The output of the feedline carrying signal scattered from the readout resonator is connected to the output line of the dilution refrigerator (see figure D1). Residual radiation form the multiplexer creates an elevated thermal population in the readout resonator, which is measured by the ac Stark shift of the qubit [9]. Coupling between the resonator (resonance at $\nu_r$ = 5.569 GHz, decay rate: $\kappa$ = 0.70 MHz) and qubit (ground-to-excited transition frequency at $\nu_q$ = 6.580 GHz, anharmonicity $E_c$ = 200 MHz), leads to a resonator dispersive shift of -8.5 MHz (figure D1) and an effective $\chi$-shift of -2.0 MHz.

Although the $\chi$-shift is larger than the spectroscopic linewidth of the resonator $\kappa$, we do not resolve photon number states with the two-tone spectroscopy [10]. This is a consequence of a large spectroscopic line width of the qubit line (full width at half maximum, $\Delta \nu_q \approx$ 20 MHz), which we attribute to the qubit frequency fluctuations indicated by a gaussian-like line shape (see figure D1). In the two-tone spectroscopy both readout signal and qubit excitation powers were set to low values that do not cause the ac Stark shift nor increase of the qubit's spectroscopic linewidth. As a function of supply voltage $V_{dd}$, qubit transition frequency exhibits the ac Stark shift to lower frequencies and reaches $\Delta_{ac}$ = -7.7 ± 0.3 MHz at the operating supply voltage $V_{dd}$ = 0.9 V. The direction of the ac Stark shift is in agreement with the negative $\chi$-shift and it qualitatively follows $V_{dd}$ dependent dissipated power as shown in panel (b) of figure A3.

We estimate the average thermal photon population in the readout resonator by comparing the measured $\Delta_{ac}$ with the weighted average of the expected photon number peak positions [9]. Peak positions at $2\chi i$ are weighted by the corresponding Boltzmann factor,

$$\Delta_{ac} = \frac{\sum_{i=0} 2\chi i e^{-ih\nu_r/k_B T}}{\sum_{i=0} e^{-ih\nu_r/k_B T}},$$

where $i$ is a peak number starting with 0, $h$ is the Planck constant, $k_B$ is the Boltzmann constant and $T$ is an effective electron temperature. We note that due to low $\kappa/\chi \approx$ 0.1 ratio corrections to the peak positions and the weighting factor are negligible [9]. At $V_{dd}$ = 0.9 V the ac Stark shift of $\Delta_{ac}$ = -7.7 MHz corresponds to the electron temperature of $T$ = 0.83 K and the averaged photon number of $n$ = 2.2.

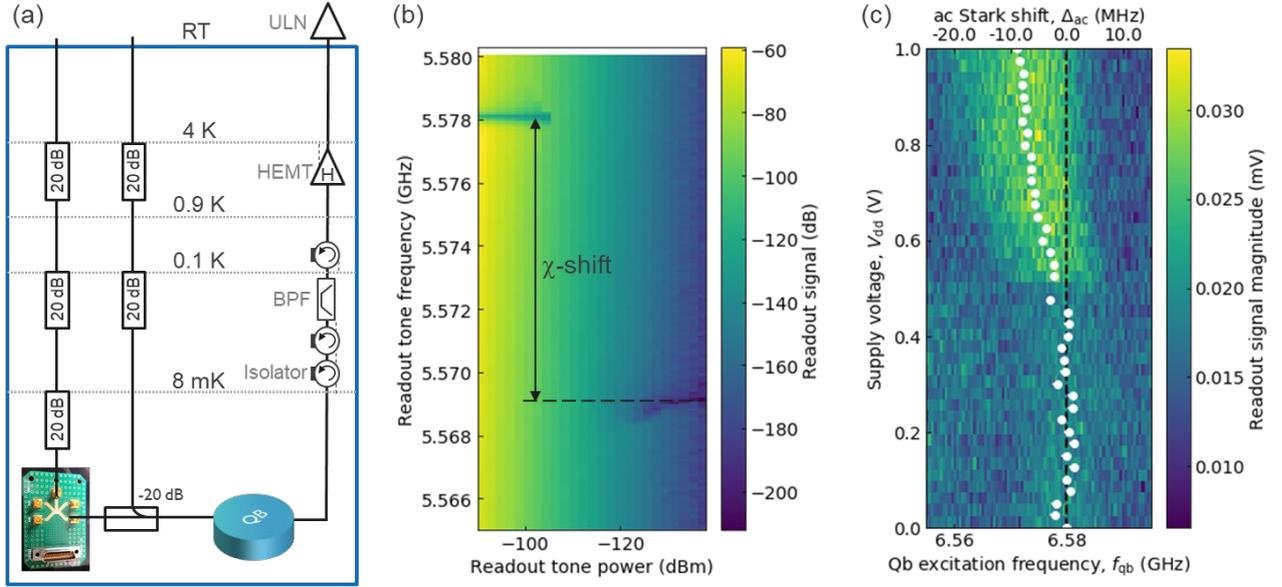

**Figure D1**: Mean thermal photon occupation in a resonator measured with a Transmon qubit. (a) Experimental setup used to measure ac Stark shift. (b) Transmission through the feedline as a function of readout tone power. Readout resonator resonance is red shifted below -110 dBm, due to the coupling to a qubit. (c) Two-tone spectroscopy measurement of g-e qubit transition frequency as a function of the cryo-CMOS multiplexer's supply voltage $V_{dd}$. Center position of approximately gaussian line shape is denoted with white dots. Shift from the reference frequency (dashed line) is attributed to the ac Stark shift. Signal change near $V_{dd} = 0.5$ V coincides with the onset of the multiplexer's operation start. This causes an impedance change in the input line system and translates to a change in the signal background. Qubit transition frequency and its linewidth are not affected by the impedance change.